\documentclass[final,3p,times,twocolumn]{elsarticle}

\usepackage{graphicx}

\usepackage{amssymb}

\RequirePackage{lineno}

\journal{Nuclear Physics B}

\begin{document}
\begin{frontmatter}


\author{D. Brandt \fnref{slac}}
\author{R. Agnese \fnref{uof}}
\author{P. Redl \fnref{stanford}}
\author{K. Schneck \fnref{slac}}
\author{M. Asai \fnref{slac}}
\author{M. Kelsey \fnref{slac}}
\author{D. Faiez \fnref{ucb}}
\author{E. Bagli \fnref{ferrara}}
\author{B. Cabrera \fnref{stanford}}
\author{R. Partridge \fnref{slac}}
\author{T. Saab \fnref{uof}}
\author{B. Sadoulet \fnref{ucb}}

 \fntext[slac]{SLAC National Accelerator Laboratory, Menlo Park, CA, USA}
 \fntext[uof]{University of Florida, Gainesville, FL, USA}
 \fntext[stanford]{Stanford University, Stanford, CA, USA}
 \fntext[ucb]{University of California Berkeley, Berkeley, CA, USA}
 \fntext[ferrara]{University of Ferrara, Ferrara, Italy} 
 \fntext[MIT]{Massachusetts Institute of Technology, Cambridge, MA, USA}

\title{Semiconductor phonon and charge transport Monte Carlo simulation using Geant4}


\author{}

\address{}

\begin{abstract}

A phonon and charge transport simulation based on the Geant4 Monte Carlo toolkit is presented. The transport code is capable of propagating acoustic phonons, electrons and holes in cryogenic crystals. Anisotropic phonon propagation, oblique carrier propagation and phonon emission by accelerated carriers are all taken into account. The simulation successfully reproduces theoretical predictions and experimental observations such as phonon caustics, heat pulse propagation times and mean carrier drift velocities.

Implementation of the transport code using the Geant4 toolkit ensures availability to the wider scientific community.

\end{abstract}


\end{frontmatter}




\section{Introduction}
\label{sec:Introduction}

We present a Monte Carlo simulation of phonon and charge transport in
semiconductor crystals using Geant4. Geant4 is a sophisticated C++ based Monte
Carlo simulation toolkit maintained by an international collaboration and
freely available under an open source license \cite{Geant-A,Geant-B}.
The toolkit was originally developed in support of High Energy Physics (HEP)
experiments and provides a framework for simulating the passage of
particles through complex geometries and materials. It aims to accurately
simulate all particle-matter interactions and has become an important tool both
for HEP particle accelerator based experiments 
, and experiments wishing to estimate backgrounds induced by comic rays and from radiogenic sources \cite{CDMS-D,Brandt}.

In its current incarnation, the Geant4 toolkit is entirely focused on free
particles and does not take into account crystal physics or conduction/valence
band interactions of the low-energy charge carriers and phonons relevant to
condensed-matter physics. This paper documents our effort to build a cohesive
Geant4 Condensed Matter Physics Monte Carlo simulation toolkit, G4CMP. The
original purpose of this project was to accurately reproduce data from the
Cryogenic Dark Matter Search (CDMS)
\cite{CDMS-A,CDMS-B,CDMS-C}, a dark-matter direct-detection experiment. The CDMS detectors are cylindrical Ge crystals
approximately $75$~mm in diameter with a height of $25$~mm \cite{CDMS-E},
cooled to approximately $60$~mK. Dark-matter particles may recoil from Ge nuclei and thus create phonons and electron-hole pairs within the crystal
\cite{Lindhart}. Electron-hole pairs drift in a few V/cm field to the crystal faces where 
they are collected. Phonons are detected by superconducting
Transition Edge Sensors (TES) \cite{TES}. We reproduce all of these
processes in our simulation. 
The phonon and charge transport code described below models several physics
processes relevant to phonon and charge collection at cryogenic
temperatures. This includes anisotropic phonon transport and focusing,
phonon isotope scattering, anharmonic downconversion, oblique charge carrier
propagation with inter-valley scattering, and emission of
Luke-Neganov phonons by accelerated carriers. The resulting G4CMP framework is
sufficiently general that it should be useful to other experiments employing
cryogenic phonon and/or ionization detectors.

\section{Phonon Transport}
\label{sec:PhononTransport}

Phonon transport was the first component of the G4CMP framework to be developed
\cite{Brandt}. Since the phonon transport code described here is intended for
temperatures $T<1 K$, scattering off thermally excited background phonons is
ignored. Currently, only acoustic phonons are simulated, with optical 
phonons not supported.

\subsection{Anisotropic transport and phonon focusing}
\label{sec:Focusing}

Phonons are quantized vibrations of the crystal lattice. The propagation of phonons is governed by the three-dimensional wave equation \cite{Wolfe}:
\begin{equation}
\label{eq:3DWave}
\rho \omega ^2e_i=C_{ijml}k_jk_me_l,
\end{equation}
where $\rho$ is the crystal mass density, $\omega$ is the angular phonon frequency, $\vec{e}$ is the polarization vector, $\vec{k}$ is a wave vector and $C_{ijml}$ is the elasticity tensor.

For any given wave vector, $\vec{k}$, Eq. \ref{eq:3DWave} has three
eigenvalues, $\omega$, and three eigenvectors, $\vec{e}$. These correspond to
the three different polarization states: Longitudinal (L), Fast Transverse (FT) and Slow
Transverse (ST). The actual direction and
velocity of propagation of phonons is given by the group velocity vector
$\vec{v}_g = d\omega/dk$. The group velocity can be calculated by interpreting $\omega$ in Eq. \ref{eq:3DWave} as a function of $\vec{k}$:

\begin{equation}
\label{eq:GroupV}
\vec{v}_g=\nabla \omega (\vec{k}),
\end{equation}

\begin{figure}
	\centering
		\includegraphics[width=0.5\textwidth]{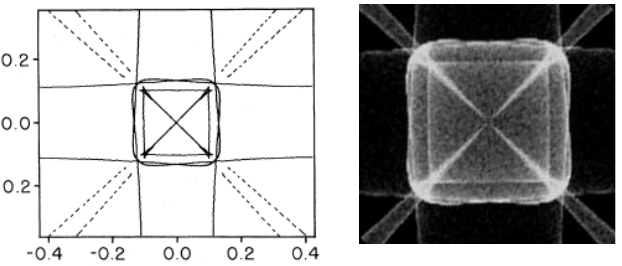}
	\caption{\textbf{Left:} outline of phonon caustics in Ge as predicted by
    Nothrop and Wolfe \cite{Nothrop}. \textbf{Right:} Phonon caustics as
    simulated using the Geant4 phonon transport code. This result is in good
    agreement with both the theoretical prediction and experimental
    observations reported by Nothrop and Wolfe \cite{Nothrop}. }
	\label{fig:caustics}
\end{figure}

Due to the anisotropy in $C_{ijml}$, Eq. \ref{eq:GroupV} yields a group
velocity $\vec{v}_g$ that is not parallel to the phonon momentum
$\hbar\vec{k}$. Instead, phonons are focused onto propagation directions that
correspond to the highest density of eigenvectors $\vec{k}$. This focusing
gives rise to caustics when observing the energy  distribution resulting from a
point-like phonon source that is isotropic in $\vec{k}$-space. The resulting caustics can be observed using micro-calorimeters \cite{Nothrop}. Figure~\ref{fig:caustics} shows that the caustics simulated by the Geant4 phonon transport code are in good agreement with experimental observations. 

For the purposes of the G4CMP phonon transport code, the wave equation is not
solved in real time. Instead, a look-up table is generated which maps
$\vec{k}$ onto $\vec{v}_g$, and bilinear interpolation is used to generate a continuous mapping function. Phonon focusing and methods for solving the three-dimensional wave equations are treated in \cite{Wolfe}. 

\subsection{Phonon processes}
\label{sec:Processes}

In addition to the phonon equation of motion, Eq.~\ref{eq:3DWave}, two processes are relevant to
acoustic phonon transport in cryogenic crystals: isotope scattering and
anharmonic downconversion \cite{Tamura1,Tamura2,Tamura3}. The scattering and
downconversion rates for phonons in a
cryogenic crystal are given by \cite{Brandt, Tamura2}:
\begin{equation}
\label{eq:anhRate}
\Gamma_{\mathrm{anh}} = A\nu^5,
\end{equation}

\begin{equation}
\label{eq:ScatterRate}
\Gamma_{\mathrm{scatter}} = B\nu^4,
\end{equation}
where $\Gamma_{\mathrm{anh}}$ is the number of anharmonic downconversion events per unit time,   
$\Gamma_{\mathrm{scatter}}$ is the number of scattering events per unit time,
$\nu$ is the phonon frequency and $A$, $B$ are constants of proportionality
related to the elasticity tensor. In Ge, $A = 6.43 \times 10^{-55} s^4$ and
$B = 3.67 \times 10^{-41} s^3$ \cite{Brandt,Tamura1}.

Isotope scattering occurs when a phonon interacts with an isotopic substitution
site in the lattice. It is effectively an elastic scattering during which the
phonon momentum vector is randomized, and the phonon polarization state can
change freely between the three states $L$, $ST$, $FT$. The partition between
the three polarizations is determined by the relative density of allowed states
for each polarization. This change between polarization states is often referred
to as \emph{mode mixing}.

Anharmonic downconversion causes a single phonon to decay into two phonons of reduced energy. This process conserves energy but not momentum, since momentum is exchanged with the crystal lattice. In theory all three polarization states can decay; however, the downconversion rate of $L$-phonons completely dominates the energy evolution of the phonon system, with downconversion events from other polarization states being negligible \cite{Tamura2}.

Equations \ref{eq:anhRate} and \ref{eq:ScatterRate} indicate that both process rates
strongly depend on phonon energy $\hbar \nu$. High-energy phonons ($\nu$ of
order THz) start out in a diffusive regime with high isotope scattering and
downconversion rates and mean free paths of order microns. Once a few
downconversion events have occurred, phonon mean free paths increase to be of
order $\sim0.1$~m (the size of a typical CDMS detector). This transition from a
diffuse to a ballistic transport mode is commonly referred to as
``quasi-diffuse'' and controls the time evolution of phonon pulses.
Simulation of phonon pulses with our Geant4 transport code as described in
\cite{Brandt} shows good agreement with experiment. Anharmonic
downconversion and isotope scattering are well understood and are discussed in
great detail in the literature \cite{Tamura1,Tamura2,Wolfe,Tamura3}.

\section{Charge Transport}
\label{sec:ChargeTransport}

In addition to the phonon physics described above, the G4CMP framework enables simulation of charge propagation through germanium crystals, for which there are three processes to consider: acceleration by an applied electromagnetic field, inter-valley scattering, and emission of Neganov-Luke phonons.

\subsection{Neganov-Luke Phonons}
As the charge carriers are accelerated through the crystal, they emit phonons
in a process that is analogous to Cerenkov radiation. Because electrons and
holes have different effective mass properties in germanium, Neganov-Luke
phonon emission is implemented differently for each type of charge carrier, as
explained below.
\subsubsection{Holes}
The effective mass of a hole in germanium is a scalar, so its
propagation is simply that of a charged particle in vacuum with an applied
field.

\begin{figure}[htpb]
    \centering
    \includegraphics[width=.5\textwidth]{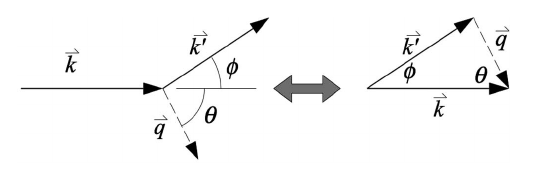}
    \caption{A charge carrier has initial wavevector $\vec{k}$ and emits a
    Neganov-Luke phonon with wavevector $\vec{q}$ \cite{Leman}}.
    \label{fig:scatter}
\end{figure}

Hole-phonon scattering is an elastic process, conserving energy and
momentum, as shown in Fig. \ref{fig:scatter}.
From conservation of energy and momentum, 
\begin{equation}
    k'^2 = k^2 + q^2 - 2kq\cos{\theta},\\
    q = 2(k\cos{\theta} - k_L),
\end{equation}
with $k_L$ defined as $k_L = mv_L/\hbar$, where $v_L$ is the longitudinal
phonon phase speed and $m$ is the effective mass of the hole \cite{Cabrera}.
Solving for $\phi$,
\begin{equation}
    \cos{\phi} = \frac{k^2 - 2k_L(k\cos{\theta} - k_L) - 2(k\cos{\theta} -
k_L)^2}{k\sqrt{k^2 - 4k_L(k\cos{\theta} - k_L})},
    \label{eq:scatterangle}
\end{equation}
where $k_L = mv_L/\hbar$. Using Fermi's Golden Rule, we can determine a
scattering rate \cite{Leman}
\begin{equation}
    1/\tau = \frac{v_Lk}{3l_0k_L}\left(1-\frac{k_L}{k}\right)^3,
    \label{eq:rate}
\end{equation}
with an angular distribution,
\begin{equation}
    P(k,\theta) d\theta =
    \frac{v_L}{l_0}\left(\frac{k}{k_L}\right)^2\left(\cos{\theta}-\frac{k_L}{k}\right)^2\sin{\theta}d\theta,
    \label{eq:ang-dist}
\end{equation}
where $0\le\theta\le\arccos{k_L/k}<\pi/2$ and $l_0$ is a characteristic
scattering length defined as $l_0 = \frac{\pi\hbar^4\rho}{2m^3C^2}$ with C
being the deformation potential constant for Ge \cite{Leman}.

\subsubsection{Electrons}
Unlike the hole, the electron has a tensor effective mass in germanium. Some
transformations need to be applied before propagating the electron through the
crystal.
\begin{figure}[htpb]
    \centering
    \includegraphics[width=.5\textwidth]{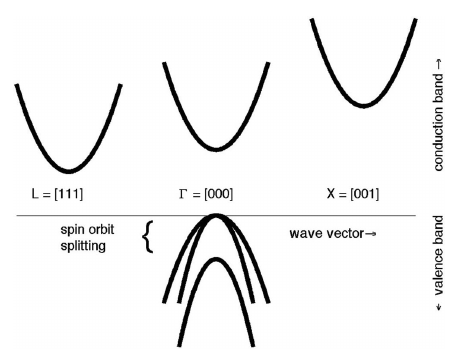}
    \caption{Conduction and valence bands in Ge. At present, the G4CMP
    framework only simulates electrons propagating in the direction of the L conduction
band. At sufficiently low temperature and applied field, this approximation is
accurate \cite{Leman}.}
    \label{fig:gebands}
\end{figure}

For a coordinate system with one axis aligned with the principal axis of the
conduction valley, the electron's equation of motion is
\begin{equation}
    \frac{eE_i}{m_i} = \frac{dv_i}{dt}.
    \label{el_eq_mtn}
\end{equation}
However, so that we can apply the same Neganov-Luke phonon emission recipe to electrons
as holes, we need to transform to a coordinate
system in which the constant energy surfaces are spherical. In that space,
$v_i^* = v_i/\sqrt{m_c/m_i}$, where $m_c$ is given by $3/m_c = 1/m_\parallel +
2/m_\perp$. And so,

\begin{equation}
    \frac{eE^*_i}{m_c} = \frac{dv_i^*}{dt},
    \label{el_eq_mtn1}
\end{equation}

Once the coordinate system is rotated into the conduction valley frame, a
Herring-Vogt transformation is applied,
\begin{equation}
    T_{HV} = \left( \begin{array}{ccc}
                    \sqrt{\frac{m_c}{m_{\parallel}}} & 0 & 0 \\
                    0 & \sqrt{\frac{m_c}{m_{\perp}}} & 0 \\
                    0 & 0 & \sqrt{\frac{m_c}{m_{\perp}}}\end{array}\right),
 \label{HV}
\end{equation}

From this space, the same recipe that applied to holes for 
Neganov-Luke phonon emission can be followed for electrons \cite{Leman}.

\subsection{Inter-Valley Scattering}
\label{sec:InterValley}

\begin{figure*}
	\centering
		\includegraphics[width=0.9\textwidth]{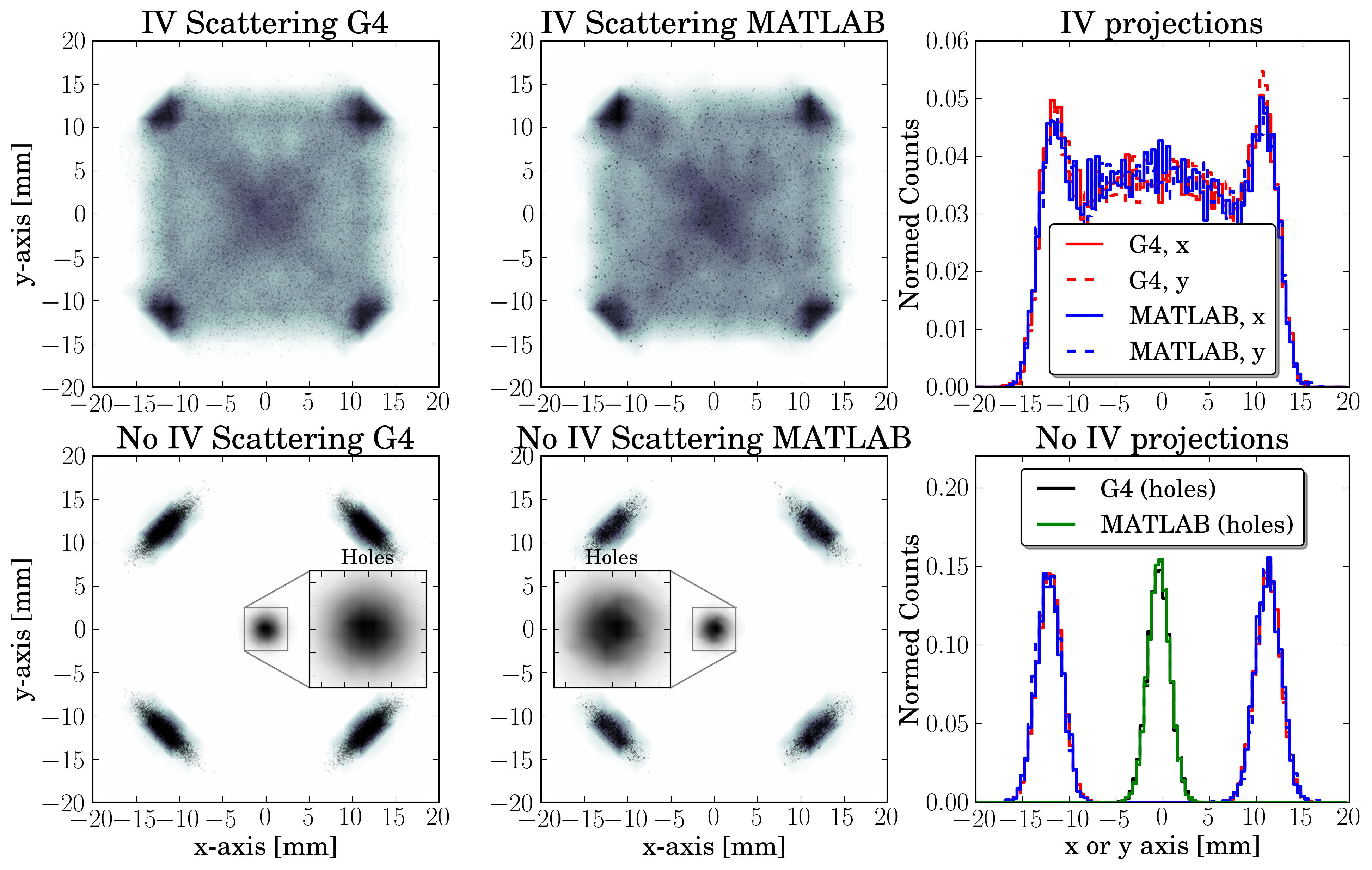}
	\caption{\textbf{Left:} Geant4 simulations \textbf{Right:} legacy MATLAB simulations. \textbf{Top Row:} Simulations with inter-valley scattering turned on, \textbf{Bottom Row:} Simulations with inter-valley scattering turned off, including hole transport through the crystal.}
	\label{fig:intervalley}
\end{figure*}

Electron propagation, as discussed in the previous section, has one
particularly interesting feature that electrons propagate through the crystal
in one of four distinct valleys \cite{Leman,Cabrera}. Electrons are not bound
to those valleys permanently, however, and can scatter between valleys. This
process is known as inter-valley scattering and occurs in one of two ways: an electron scatters off the lattice or off an impurity \cite{iv}. The rate for both processes is dependent on the electric
field strength, with lattice scattering being the dominant factor in larger
fields ($\gtrsim$5~V/m), while impurity scattering dominates in lower fields
($\sim$1~V/m). The EDELWEISS \cite{edelweiss} collaboration determined the
scattering rates as a function of the electric field for typical Ge crystals
\cite{iv}. We used the results obtained in these studies to set the inter-valley scattering amplitude in the Geant4 framework and compare it to previous implementations of the charge transport 
code \cite{Leman,Cabrera}. The result of electrons and holes propagating
through 2.54~cm of Ge in an 0.5~V/m electric field is shown in
Fig. \ref{fig:intervalley}. The top two panels show the result with
inter-valley scattering turned on while the bottom two panels show the result
for inter-valley scattering turned off. The panels on the right show the
results for the legacy (MATLAB-based) simulation \cite{Leman,Cabrera} with
somewhat fewer statistics than the Geant4 simulations. The bottom two panels
also show hole transport through the crystal for the Geant4 and legacy
simulations \cite{Leman,Cabrera}.

\subsection{Charge Carrier Drift Speed Results}
\begin{figure}[htpb]
    \centering
    \includegraphics[width=.43\textwidth]{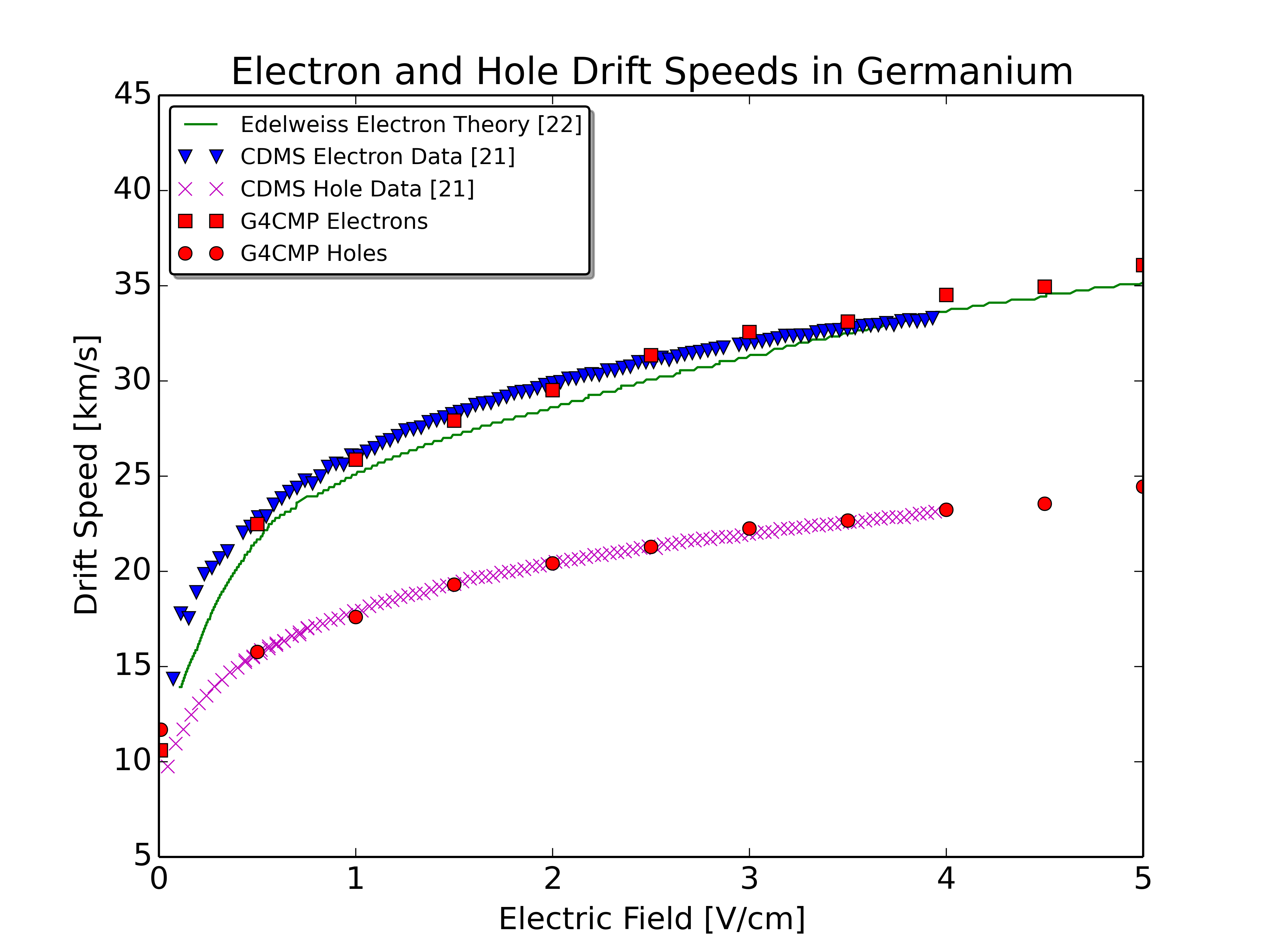}
    \caption{Comparison of the G4CMP charge code with various data and models.}
    \label{fig:results}
\end{figure}

As the charge carriers are accelerated by the electric field, they emit
Neganov-Luke
phonons. For different values of the applied electric field, a maximum drift
speed of the carrier is reached, where the energy gained from the electric
field is canceled by the energy lost from emission of phonons. Fig.
\ref{fig:results} shows the average drift velocities of electrons and holes in
the G4CMP package with experimental data \cite{Sundqvist} and the theoretical curve \cite{Fortuna} on which the
Monte Carlo is based. 

\section{Conclusion}
\label{sec:Conclusion}


We have presented a new Monte Carlo transport code capable of propagating phonons on a
cryogenic crystal lattice (Section \ref{sec:PhononTransport}) as well as
drifting electron-hole pairs, taking into account conduction band anisotropy
(Section \ref{sec:ChargeTransport}). The results produced by the transport code
are in good agreement with experiment, reproducing phonon caustics and carrier
drift velocity with acceptable accuracy. It was shown that this code reproduces
heat-pulse propagation and dispersion in cryogenic Ge crystals with acceptable
accuracy in \cite{Brandt}. The entire transport code is written in C++ for Geant4 and
is easily adaptable for crystals other than Ge, provided that the Herring-Vogt contracted elasticity tensor and effective carrier masses are known. The Geant4 framework makes it possible to extend the code presented here. The phonon transport code is already freely available as part of the examples provided with Geant4 v9.6p02 and newer. 
We hope that the work presented here will establish Geant4 as a tool in
condensed-matter physics and cryogenic calorimeter design as well as motivate others to add to the G4CMP framework.
\newline
\newline
We would like to thank members of the CDMS Detector Monte Carlo Group for discussions that have made completing this work 
possible. Specifically we want to thank P.L. Brink, A. Anderson and C. Schlupf. Furthermore we would like to thank A. Phipps and K. Sundqvist for 
useful discussions and providing us with experimental data. S. Yellin and R. Bunkers' on point feedback on earlier paper drafts have helped us 
improve the final paper tremendously. 
This work is supported in part by the National Science Foundation and by the United States Department of Energy. SLAC is operated under Contract No. DE-AC02-76SF00515 with
the United States Department of Energy.










\end{document}